\documentclass[aps,prd,showpacs,preprintnumbers,superscriptaddress,nofootinbib]{revtex4}

 \usepackage{graphicx}
 \usepackage{enumerate}
\usepackage{multirow}

\usepackage{hyperref}

\def\Journal#1#2#3#4{{#1} {{\bf #2},} {#4} {(#3)}}

\def\PLB{{Phys. Lett.}  B}

\def\PRP{{ Phys. Rep.}}

\def\PR{{Phys. Rev.}}
\def\PRD{{Phys. Rev.} D}
\def\PRC{{Phys. Rev.} C}

\def\ZPC{{Z. Phys. C}}

\def\EPJC{{Eur. Phys. J.} C}

\def\MPLA{{Mod. Phys. Lett.} A}

\def\JHEP{{J. High Energy Phys.}}

\def\FBS{Few Body Syst.}
\def\IJM{Int. J. Mod. Phys.}

\def\ra{\rightarrow}

\def\be{\begin{equation}}
\def\ee{\end{equation}}
\def\bea{\begin{eqnarray}}
\def\eea{\end{eqnarray}}

\def\qbar{{\bar q}}

\usepackage{relsize}
\def\babar{\mbox{\slshape B\kern-0.1em{\smaller A}\kern-0.1em
    B\kern-0.1em{\smaller A\kern-0.2em R}}}

\begin{document}

\title{Radial excitation of light mesons and the $\gamma \gamma^* \rightarrow \pi^0$ transition form factor}

\author{F.-G. Cao}
\affiliation{School of Fundamental Sciences PN461, Massey University, \\ Private Bag 11 222,
Palmerston North, New Zealand}

\date{\today}
\begin{abstract}

We investigate radial excitation of the quark-antiquark pair in the $\pi^0$ meson
and its effects on the $\gamma \gamma^* \ra \pi^0$ transition form factor in the framework of light-cone perturbative QCD.
The existing constraints on the light-cone wave function of the lowest Fock state $|q \qbar \rangle$  in the $\pi^0$ meson allow 
a sizeable radial excitation of the quark-antiquark pair. 
We construct the light-cone wave function for the quark-antiquark pair in the first radially excited state (the 2S state) 
using a simple harmonic oscillator potential. 
The distribution amplitude obtained for the 2S state has two nodes in $x$ at low scale of $Q$ and thereby has a much strong 
scale dependence than the 1S state. 
Contributions from this radial excitation to the  $\gamma \gamma^* \ra \pi^0$ 
transition form factor exhibit different $Q^2$-dependence behavior from the ground state and thus can modify the prediction for the
transition form factor in the medium-large region of $Q^2$.
\end{abstract}


\maketitle

\section{Introduction}
Light-cone perturbative QCD has been applied to the calculations for many inclusive and exclusive hadronic processes. 
The test of these calculations against available experimental data, particularly for many exclusive processes, 
is usually retarded by the possible higher order and higher twist contributions 
to the theoretical calculations at low and medium regions of momentum transfer ($Q^2$) and the limited availability of experimental data at high $Q^2$.
The measurements of the $\gamma \gamma^* \ra \pi^0$ transition form factor (TFF) \cite{BaBar2009,Belle2012}, the simplest QCD process involving a 
hadron, posed a very interesting challenge to the theoretical calculations. While the results from the BaBar Collaboration \cite{BaBar2009}
show a rapid growth for $Q^2 > 15~{\rm GeV}^2$ and a significant deviation from the asymptotic prediction from perturbative QCD, 
the results from the Belle Collaboration \cite{Belle2012}
are much more in agreement with theoretical expectations.
A lot of theoretical studies have been done
(see, for example, \cite{LiM2009,MikhailovS2009,WuH2010,RobertsRBGT2010,BroniowshiA2010,PhamP2011,Kroll2010,
BrodskyCTe2011_1,BrodskyCTe2011_2,BalakirevaLM2012,LhiZG2012,StefainsBMP2012,LiSW2014,MeloEF2014,MikhailovPS2014,
EichmannFWW2017,ChoiRJ2017,Stefanis2020,ZhongZFWH2021})
in explaining these measurements for the pion TFF.

In this work, we investigate the possible radial excitation of the quark-antiquark ($q\qbar$) pair in the $\pi^0$
and study its contribution to the $\gamma \gamma^* \ra \pi^0$ transition form factor. 
The excited $q\qbar$ pair, when in the first radially excited state with the lowest angular momentum (i.e. the 2S state),
has the same quantum numbers as the ground state (the 1S state), i.e. $J^{P C}=0^{- +}$; 
such an excitation does not violate any fundamental principles of QCD and thus is allowed in the quark models of hadrons.
In Section \ref{Sec:WFConstraints}, we construct the light-cone wave function for the $q \qbar$ pair in the 2S state 
using the wave function in the center-of-mass (CM) frame for the simple harmonic
oscillator potential, and the Brodsky, Huang and Lepage (BHL) prescription \cite{BHL,GLepageBHM81} for connecting the equal-time (instant-form) 
wave function in the centre-of-mass frame and the light-front wave function.
We exam the existing constraints on the pion light-cone wave function and the possibility for the $ q \qbar$ pair in the 2S state. 
In Section \ref{Sec:PionTFF}, we calculate the contribution of the possible 2S state to the  $\gamma \gamma^* \ra \pi^0$ transition form factor.
A summary is given in the last Section.

\section{Light-cone wave function and distribution amplitude with radial excitation}
\label{Sec:WFConstraints}

Light-cone wave functions (LCWFs) are universal and contain all nonperturbative
information of partons in the hadrons. 
The light-cone wave functions can be deduced from experimental data and/or
from nonperturbative QCD computations \cite{SBrodskyPP98}.
In this study we adopt a widely used method which was first suggested by Brodsky, Huang and Lepage \cite{BHL,GLepageBHM81}
in computing the LCWFs for the mesons.
In this method, a connection between the equal-time (instant-form) wave function in the centre-of-mass (CM)
frame, usually computed using effective potentials for the $q \qbar$ pair,
and the light-front wave function is established by equating the off-shell propagator 
$\varepsilon=M^2-{({\sum}_{i=1}^nk_i)}^2$ in the two frames,
\begin{displaymath}
\varepsilon=\left\{\begin{array}{ll} M^2-(\sum_{i=1}^n q_i^0)^2,
&\sum_{i=1}^n\textbf{q}_i=0, ~{\mathrm{[CM]}},\\
M^2-\sum_{i=1}^n \frac{{\bf k}_{\perp i}^2+m_i^2}{x_i},
&\sum_{i=1}^n {\bf k}_{\perp i}=0,~ \sum_{i=1}^n x_i=1~ {\mathrm{[LC]}},
\end{array}\right.
\end{displaymath}
where ${\bf q}_i$ is the momentum of constituent~$i$ in the CM frame,
and $x_i$ and ${\bf k}_{\perp i}$ are the light-cone momentum fraction and transverse momentum of the constituent~$i$
in the light-cone frame.
For a two-identical-particle system with $m \equiv  m_1=m_2$, ${\bf q} \equiv {\bf q_1}=-{\bf q_2}$, 
${\bf k}_{\perp} \equiv {\bf k}_{\perp 1}=-{\bf k}_{\perp 2}$, and $x \equiv x_1=1-x_2$, one has
\begin{equation}\label{Q-square1}
\textbf{q}^2=\frac{{\bf k}^2_{\perp}+m^2}{4x(1-x)}-m^2,
\label{eq:propagator}
\end{equation}
and thereby can establish a relationship for the wave functions both in the light-cone form $\psi(x,{\bf k_\perp})$
and in the equal-time form  $\psi({\bf q})$,
\bea 
\psi \left(\frac{{\bf k}^2_{\perp}+m^2}{4x(1-x)}-m^2\right) \leftrightarrow \psi({\bf q}^2).
\label{eq:relation}
\eea

For the equal-time wave function in the CM frame,
we employed the wave function computed with the
harmonic oscillator potential for the quark-antiquark system, $V({\bf r})=\frac{1}{2} m \omega^2 {\bf r}^2$,
where ${\bf r}$ is the separation between the quark and antiquark.
The wave functions for the ground state (1S) and the first radially excited state (2S) are \cite{DFaimanH1968},
\bea
\psi^{1S}(r) &=& \left( \frac{2B}{\pi} \right)^{3/4}{\rm exp}(-B r^2), \label{eq:wf1S}\\
\psi^{2S}(r) &=&\sqrt{\frac{3}{2}} \left( \frac{2B}{\pi} \right)^{3/4} (1-\frac{4}{3}B r^2){\rm exp}(-Br^2), \label{eq:wf2S}
\eea
where $B=\frac{1}{2} m \omega$.
The wave functions in the momentum space can be written in the form of
\begin{eqnarray}
\psi^{1S}(\bf{q}) & \propto & {\rm  exp} \left(-\frac{{\bf q}^2}{4B} \right), \\
\psi^{2S}(\bf{q}) & \propto & ({\bf q}^2-3B){\rm exp} \left(-\frac{{\bf q}^2}{4B}\right).
\end{eqnarray}

Using relationship Eq.~(\ref{eq:relation}) one can obtain the spatial parts of light-cone wave functions for
the 1S and 2S states. The obtained LCWF for the ground state is the well-known Gaussian form. 
In the widely-used factorized Gaussian form for the LCWF the dependence on the quark mass $m$ 
is absorbed into the overall normalization factor for the wave function. 
Thus we use the following forms of the LCWFs for the $q \qbar$ pair in the 1S and 2S state, 
\bea
\psi^{1S}_{q \bar q}(x,\mathbf{k_\perp})&=&
 a_1  {\rm exp}\left[ - \frac{1}{4 B} \left( \frac{{\bf k}^2_{\perp}}{4x(1-x)} \right) \right] ,
 \label{eq:wf1S} \\
\psi^{2S}_{s \bar s}(x,\mathbf{k_\perp})&=&
a_2 \left( \frac{{\bf k}^2_{\perp}}{4x(1-x)} - 3B \right)
{\rm exp}\left[ - \frac{1}{4 B} \left( \frac{{\bf k}^2_{\perp}}{4x(1-x)} \right) \right],
\label{eq:wf2S}
\eea
where the parameters $a_1$, $a_2$ and $B$ can be fixed, in principle, by considering existing constraints on the pion LCWF.

\begin{figure}[h]
\begin{center}
\includegraphics[width=10.5cm]{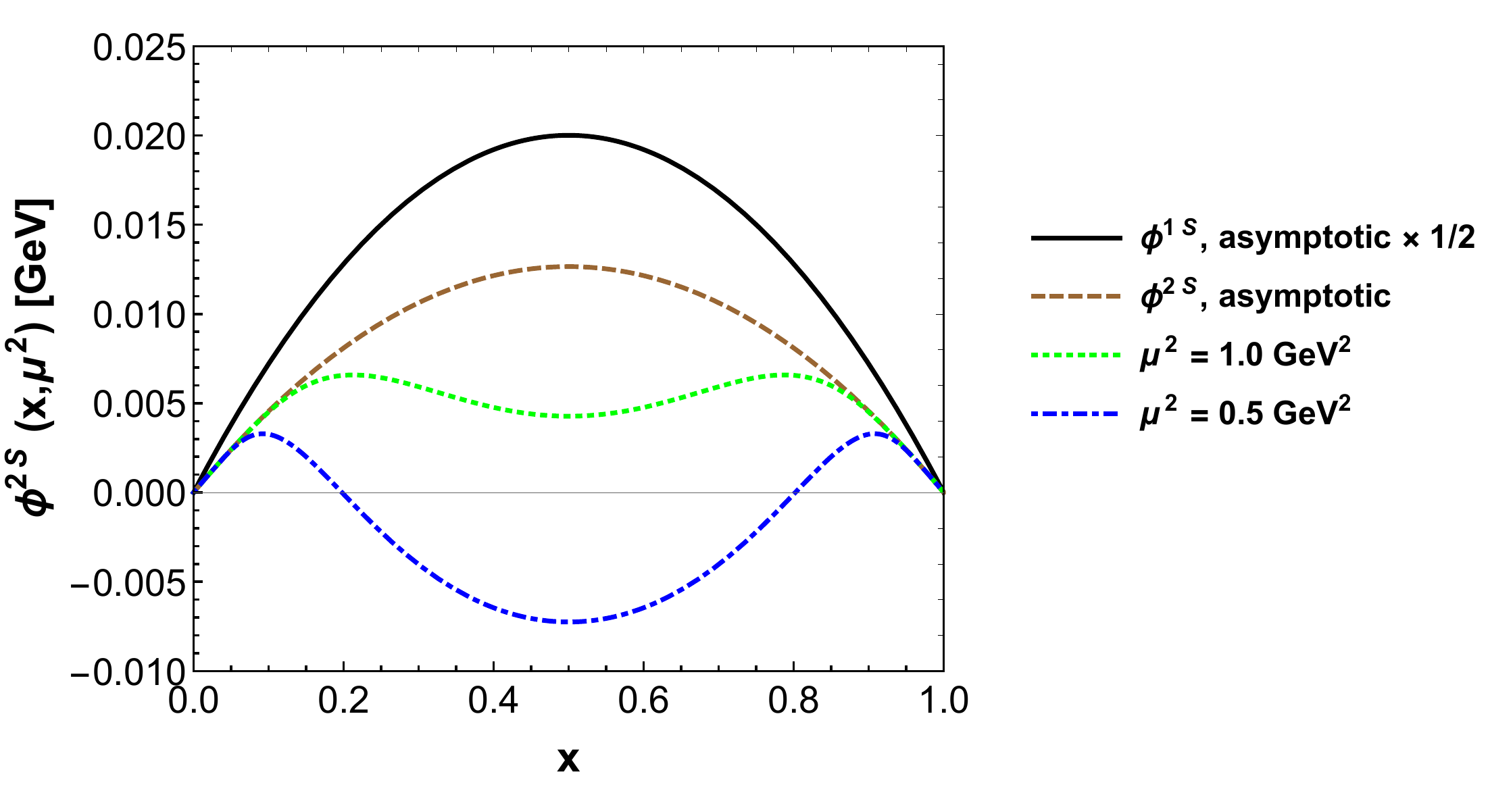}
\end{center}
\caption{\label{fig:DA2S} \small{Soft evolution of the pion distribution amplitude for the 2S state. The dot-dashed and dotted
curves (color online) are for $\mu^2=0.5$ and $1$ GeV$^2$, respectively. The dashed cure is for the distribution amplitude 
at the asymptotic limit $\mu^2 \ra \infty$.
The solid curve is for 1/2 of the asymptotic form $\phi(x)= \sqrt{3} f_\pi x(1-x)$ for the 1S state.
The parameters in the LCWF are taken to be $B=0.0843$ GeV$^2$ and $a_2=70.3$ GeV$^{-3}$.}}
\end{figure}

The pion distribution amplitude (DA) is defined in the light-front formalism as the integral 
of the valance $q \qbar$ LCWF in light-cone gauge $A^+=0$,
\bea
\phi(x,\mu)=\int^{\mu}_0 \frac{d^2\mathbf{k_\perp}}{16\pi^3} \psi_{\pi} (x,\mathbf{k_\perp}),
\eea
where $\mu$ is an arbitrary scale.
Thus one can define the distribution amplitudes for the 1S and 2S states, $\phi^{\rm 1S}(x,\mu)$ and  $\phi^{\rm 2S}(x,\mu)$,
using the LCWFs given by Eqs.~(\ref{eq:wf1S}) and (\ref{eq:wf2S}),
\bea
\phi^{1S}(x,\mu)&=&\frac{a_1 B} {\pi^2} x (1-x)\left( 1- {\rm exp} \left[ - \frac{\mu^2}{16 B x (1-x)}\right]\right),
\label{eq:DA1S} \\
\phi^{2S}(x,\mu)&=&\frac{a_2 B^2} { \pi^2} x(1-x) \nonumber \\
& &
\times \left(
 1 - \left[1 + \frac{\mu^2}{4 B x (1-x)} \right] {\rm exp} \left[ - \frac{\mu^2}{16 B x (1-x)} \right]
\right).
\label{eq:DA2S}
\eea
Both distribution amplitudes turn into the asymptotic form, $\phi(x) \sim x (1-x)$, at large scale of $\mu$.
However, the exponential dependence on the scale $\mu$, via the combination $-\mu^2/[4 B x(1-x)]$, 
means the DAs could be substantially different from the asymptotic form for small $\mu$. 
Furthermore, the DA for the 2S state have two nodes, with negative values at the medium $x$ region for $\mu^2 < 11.35 B$
(see Fig.~\ref{fig:DA2S}). 

Four constraints have been identified in determining the parameters in the LCWF 
of the lowest Fock state of the pion \cite{THuangMS94}.
The lepton decay of $\pi \ra \mu \nu$ suggests
\bea
\int^1_0 dx \phi(x, \mu) = \frac{f_\pi}{2 \sqrt{3}},
\label{eq:Constraint1}
\eea
where $f_\pi=92.4$ MeV is the pion decay constant.
This constraint depends on the choice of the scale $\mu$, although it is common practice in the literature to
choose a high enough value for $\mu$ in aiming to reduce this dependence on the choice of scale for any calculations. 

The second constraint is obtained \cite{BHL} by relating chiral anomaly prediction for the $\pi^0 \ra \gamma \gamma$
decay width to the light-cone formalism prediction for the TFF at the limit $Q^2 \ra 0$,
\bea
\int^1_0 d x \psi_\pi (x, 0_\perp) = \frac{\sqrt{3}}{f_\pi}.
\label{eq:Constraint2}
\eea
The prediction for the TFF at $Q^2 \ra 0$ in the light-cone formalism involves a divergent integral and relies on
the expansion of the wave function under the condition of $q_\perp \ll k_\perp$ (with $Q^2=q_\perp^2$) \cite{BHL}.
This prescription could be invalid when the 2S state is included in the calculation since the wave function for the 2S state 
given by Eq.~(\ref{eq:wf2S}) has an additional dependence on transverse momentum apart from the exponential factor. 

The third constraint is the result from requiring the 
probability of finding the $q \qbar$ Fock state in the pion to be smaller than 1,
\bea
P_{q \qbar }=\int^1_0 dx \int^\infty_0 \frac{d^2\mathbf{k_\perp}}{16\pi^3} \left| \psi_{\pi} (x,\mathbf{k_\perp}) \right| \leq 1.
\label{eq:Constraint3}
\eea
With the wave functions given by Eqs. (\ref{eq:wf1S}) and (\ref{eq:wf2S}), the probabilities are found to be
$P_{q\qbar}^{1S}=\frac{a_1^2 B}{12 \pi^2}$ and $P_{q\qbar}^{2S}=\frac{5 a_2^2 B^3}{12 \pi^2}$ for the 1S and 2S states, respectively.
The requirement of the probability being smaller than 1 puts constraints on the combination of  parameters $a_1$ and $a_2$ with $B$.

The fourth constraint comes from considerations of the average quark transverse momentum of the pion which is defined 
as the root-mean-squared for the quark transverse momentum, 
$\langle k_\perp \rangle =\sqrt{\langle \mathbf{k}_\perp^2 \rangle_{q \qbar}}$, with
\bea
\langle \mathbf{k}_\perp^2 \rangle_{q \qbar} = 
\int^1_0 dx \int^\infty_0 \frac{d^2\mathbf{k_\perp}}{16\pi^3} \left|  \mathbf{k_\perp}\right|^2 
\left| \psi_{\pi} (x,\mathbf{k_\perp}) \right|^2 /P_{q \qbar}.
\label{eq:Constraint4}
\eea
It is known experimentally that the average quark transverse momentum of the pion should be about a few hundreds MeV. 
Thus the average quark transverse momentum of the pion in the lowest Fock state should be in the range of a few hundreds MeV.
For the wave functions given by Eqs. (\ref{eq:wf1S}) and (\ref{eq:wf2S}), the average quark transverse momentum is 
proportional to $\sqrt{B}$, with $\langle k_\perp \rangle ^{1S}_{q \qbar}=\sqrt{8/5}\sqrt{B}$ and
$\langle k_\perp \rangle ^{2S}_{q \qbar}=\sqrt{72/25}\sqrt{B}$ for the 1S and 2S states, respectively.
Thus considerations of the average quark transverse momentum provide information for the parameter $B$ only.

Equations (\ref{eq:Constraint1}), (\ref{eq:Constraint2}), (\ref{eq:Constraint3}) and (\ref{eq:Constraint4})
form a set of constraints on the pion LC wave function.
Applying those constrains for the 1S state, we have
\bea
B a_1 &\geq& \sqrt{3}\pi^2 f_\pi~~~({\rm lower~limit~applicable~for~} \mu \ra \infty), \label{eq:Constraint1.1}\\
a_1&=&\frac{\sqrt{3}}{f_\pi}, \\
B a_1^2&=&12 \pi^2 P_{q\qbar}^{1S}, \\
B &=&\frac{5}{8} \left( \langle k_\perp \rangle ^{1S}_{q \qbar} \right)^2.
\eea
Requiring $\langle k_\perp \rangle ^{1S}_{q \qbar}$ to be in the range of $(300 \sim 500)$ MeV together with those constraints
we have $a_1 = {\sqrt{3}}/{f_\pi}$ and $B = (0.0843 \sim 0.156)~ {\rm GeV}^2$,
which suggests the probability $P_{q \qbar}^{1S}$ is in the range
of $0.25 \sim 0.46$\footnote{The lower limit is set by Eq.~(\ref{eq:Constraint1.1}).}.
The obtained probability
suggests the higher Fock state components such as $|q \qbar g \rangle$, $|q \qbar q \qbar \rangle$ etc
and/or the radial excitation of the lowest Fock state $|q \qbar \rangle$ in the pion are significant. While there are limited
discussions on the former possibility, we are not aware of any discussion on the latter possibility and its significance
as we mentioned in the previous section.

The parameter $a_2$ cannot be determined by using the first and the second constraints [Eq.~(\ref{eq:Constraint1})
and Eq.~(\ref{eq:Constraint2})] since the DA for the 2S state have two nodes when $\mu$ is not very large and the contribution from 
the 2S state to the integral in Eq.~(\ref{eq:Constraint1}) could vanish while it is not clear whether Eq.~(\ref{eq:Constraint2}) is
applicable for the 2S state. We will set $a_2$ to be such a value that the probability for the 2S state is 1/2 of that for the 1S state
and study its effect on the $\pi^0 \ra \gamma \gamma$ transition from factor in the next section.
Taking $P_{q\qbar}^{1S}=25\%$ and $P_{q\qbar}^{2S}=12.5\%$ we have $B=0.0843$ GeV$^2$, $a_1=18.4$ GeV$^{-1}$ and
 $a_2=70.3$ GeV$^{-3}$.
 The corresponding average quark transverse momenta for the 1S and 2S states are $367$ MeV and $493$ MeV, respectively. 

\begin{figure}[h]
\begin{center}
\includegraphics[width=10cm]{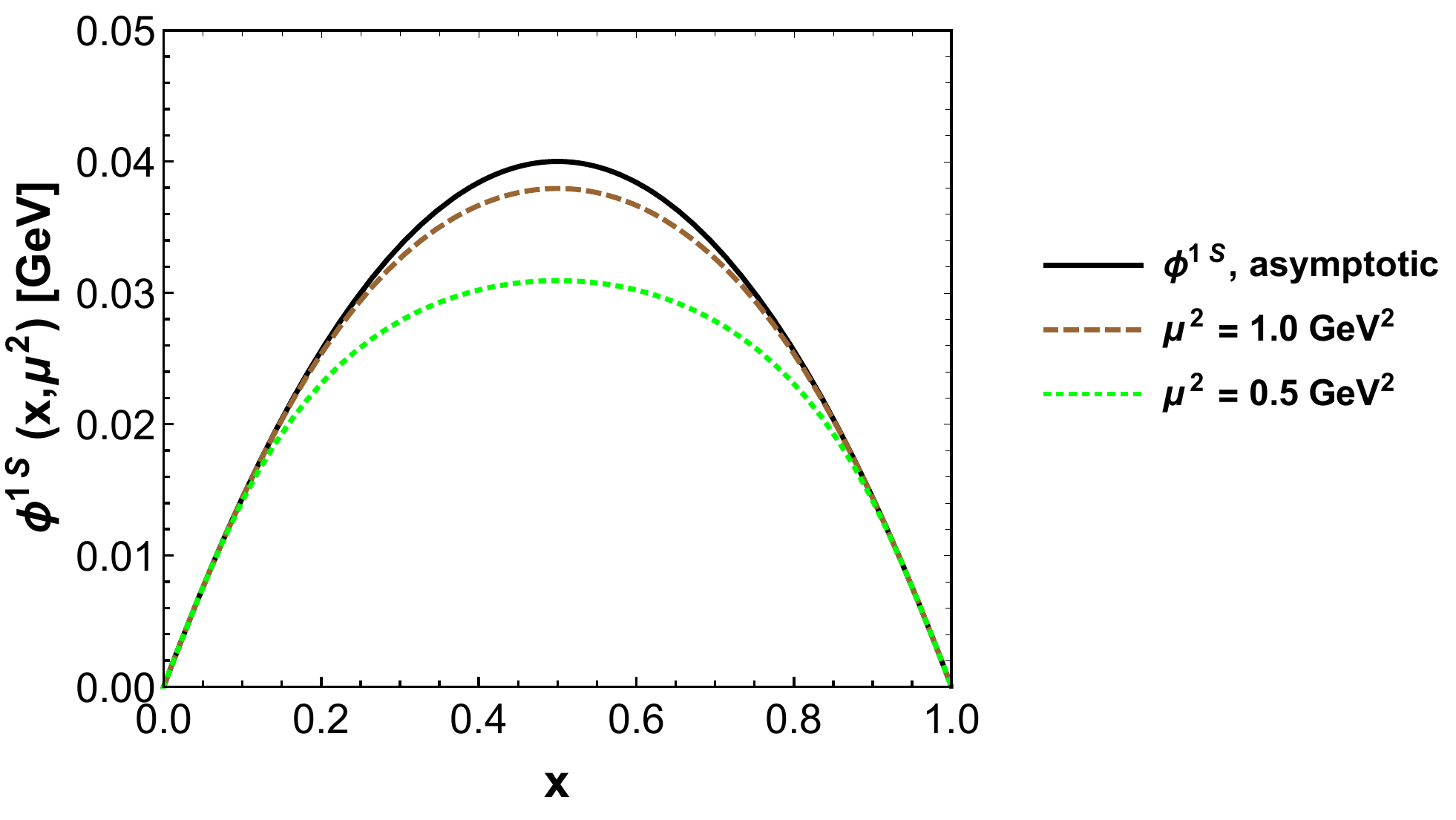}
\end{center}
\caption{\label{fig:DA1S} \small{Soft evolution of the pion distribution amplitude for the 1S state. The dotted and dashed curves (color online)
are for $\mu^2=0.5$ and $1$ GeV$^2$, respectively. The solid curve is for the asymptotic form $\phi(x)= \sqrt{3} f_\pi x(1-x)$.
The parameters in the LCWF are taken to be $B=0.0843$ GeV$^2$ and $a_1=18.4$ GeV$^{-1}$.}}
\end{figure}

The dependence of DAs on the scale $\mu$, which is the `soft' QCD evolution of the DA discussed in \cite{BrodskyCTe2011_1},
is shown in Figs.~\ref{fig:DA1S} and \ref{fig:DA2S} for the 1S and 2S sates, respectively. 
The DA for the 2S state could have negative values in the medium $x$ region for $\mu^2 < 11.35B$.
This `soft' QCD evolution has a far greater impact on the 2S state than on the 1S state. 
This difference in the scale dependence of the two DAs will modify QCD predictions for exclusive processes in a large range of 
$Q^2$ when the 2S state contributions are included.

\section{The effects on the $\gamma \gamma^* \ra \pi^0$ transition form factor}
\label{Sec:PionTFF}

The existence of radially excited components in the pion, though not at a very significant level,
may have potential impacts on various processes. 
We study the effects on the $\gamma \gamma^* \ra \pi^0$ transition form factor in the framework of light-cone perturbative QCD.
The transition form factor can be written as, 
when the involved momentum transfer $Q^2$ is large \cite{BL8081,PQCD,FCaoHM96}, 
\begin{equation}
F_{\pi\gamma}(Q^2)=2\sqrt{3}(e_u^2-e_d^2)\int^1_0d x\int
\frac{d^2\mathbf{k_\perp}}{16\pi^3} \psi_{\pi}(x,\mathbf{k_\perp})
T_H(x,\mathbf{k_\perp},Q),
\label{eq:Fpigamma}
\end{equation}
where $\psi_{\pi}(x,\mathbf{k_\perp})$ is the pion light-cone wave function and $T_H$ is
the hard scattering amplitude,
\begin{equation}
T_H(x,\mathbf{k_\perp},Q)=\frac{\mathbf{q_\perp}\cdot({\bar x}\mathbf{q_\perp}+\mathbf{k_\perp})}
{\mathbf{q_\perp^2}({\bar x}\mathbf{q_\perp}+\mathbf{k_\perp})^2} +(x \leftrightarrow {\bar x}),
\label{eq:hard}
\end{equation}
with ${\bar x}=1-x$ and $\mathbf{q_\perp^2}=-Q^2$. Calculations based
on Eq.~(\ref{eq:Fpigamma}) give a good description of experimental
data for the meson transition form factors in a large range of
$Q^2$ ($Q^2 > $ a few GeV$^2$)  \cite{FCaoHM96}.

\begin{figure}[h!]
\begin{center}
\includegraphics[width=7cm]{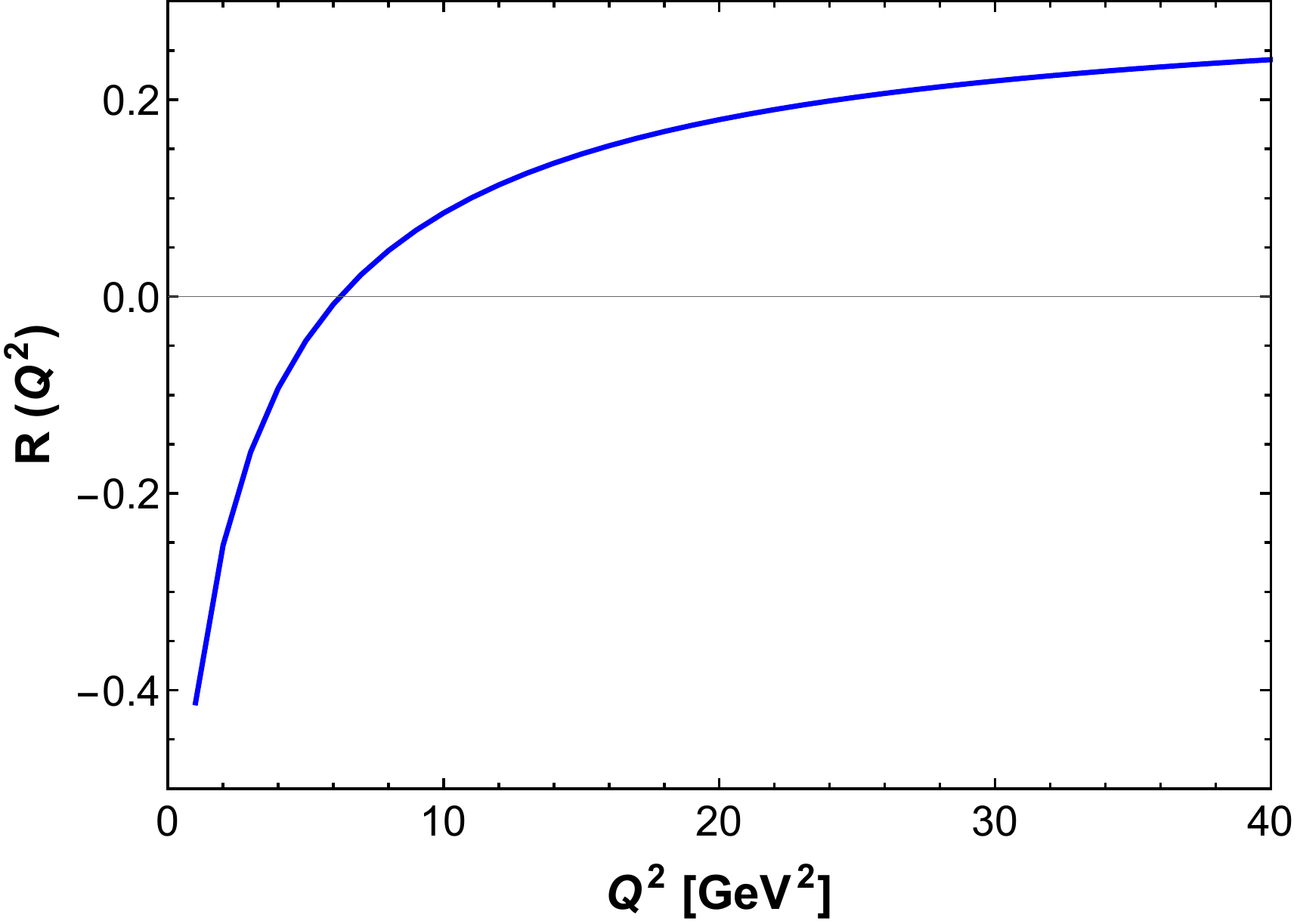}
\label{fig:Q2PionTFF}
\end{center}
\caption{\label{fig:Ratio} \small{The ratio of the contribution to the transition form factor from the 2S state to that from the 1S state,
$R(Q^2)=F^{2S}_{\pi\gamma}(Q^2)/F^{1S}_{\pi\gamma}(Q^2)$.}}
\end{figure}

The LCWFs for both the 1S and 2S states, Eqs.~(\ref{eq:wf1S})and (\ref{eq:wf2S}),
depend on the transverse momentum through $\mathbf{k}_\perp^2$. The transition form factor can be expressed in 
terms of the distribution amplitudes for the 1S and 2S sates,
\bea
Q^2 F_{\pi\gamma}(Q^2)=\frac{4}{\sqrt{3}} \int_0^1 \frac{dx}{\bar x} 
\left[ \phi^{1S}(x,{\bar x} Q) + \phi^{2S}(x,{\bar x} Q)\right].
\label{eq:Q2Fpigamma}
\eea
The first term in Eq.~(\ref{eq:Q2Fpigamma}) represents the contribution from the 1S state 
while the second term represents the contribution from the 2S state.
The ratio of the contribution to the transition form factor from the 2S state to that from the 1S state,
$R(Q^2)=F^{2S}_{\pi\gamma}(Q^2)/F^{1S}_{\pi\gamma}(Q^2)$ is shown in figure \ref{fig:Ratio}.
The contribution from the 1S state is positive for all $Q^2$. Figure \ref{fig:Ratio} shows that the contribution from the 
2S state changes sign from negative to positive at $Q^2$ just above 6 GeV$^2$. 
In terms of the magnitude, the contribution from the 2S state is about 40\% of that from the 1S state at
$Q^2=1$ Gev$^2$ and 25\% at $Q^2 = 40$ GeV$^2$.
It is the combination ${\bar x Q}$ that sets the scale in the DAs, thus the TFF at a scale $Q$ is determined
by the DAs at all scales up to $Q^2$, rather than just the DAs at that particular scale.
The 2S state DA evolves more significantly than the 1S state DA. Thus including the contribution from the 2S state 
will introduce stronger dependence on $Q^2$ than including the 1S state only for a large range of $Q^2$.
In figure \ref{fig:Q2PionTFF} we show the contributions from both the 1S and 2S states, and the total for the transition form factor,
in comparison with the data from various experimental groups.
Including the 2S state contribution in the calculations makes the TFF grow faster than including the 1S state only,
but the growth is still far below what is suggested by the BaBar data.
This reaffirms the conclusion made in \cite{RobertsRBGT2010,BrodskyCTe2011_1,BrodskyCTe2011_2} that 
the TFF measured by the BaBar Collaboration is difficult to explain in the current framework of QCD.
Including the 2S state contribution in the calculations improves the agreement with the Belle's measurement. 
The next-to-leading order corrections to the TFF are about $10\%$ at $Q^2 \sim 30$ GeV$^2$ \cite{BrodskyCTe2011_1}.
A better agreement with the Bell's data could be achieved by including the next-to-leading corrections and fine turning the 
probability for the 2S state to be in the range of $10 \sim 15 \%$.
The calculations for the small $Q^2$ region ($Q^2<10$ GeV$^2$) are much smaller than the experimental data as 
the nonperturbative contributions are expected to be dominant in this region \cite{WuH2010,ZhongZFWH2021}. 

\begin{figure}[h!]
\begin{center}
\includegraphics[width=9cm]{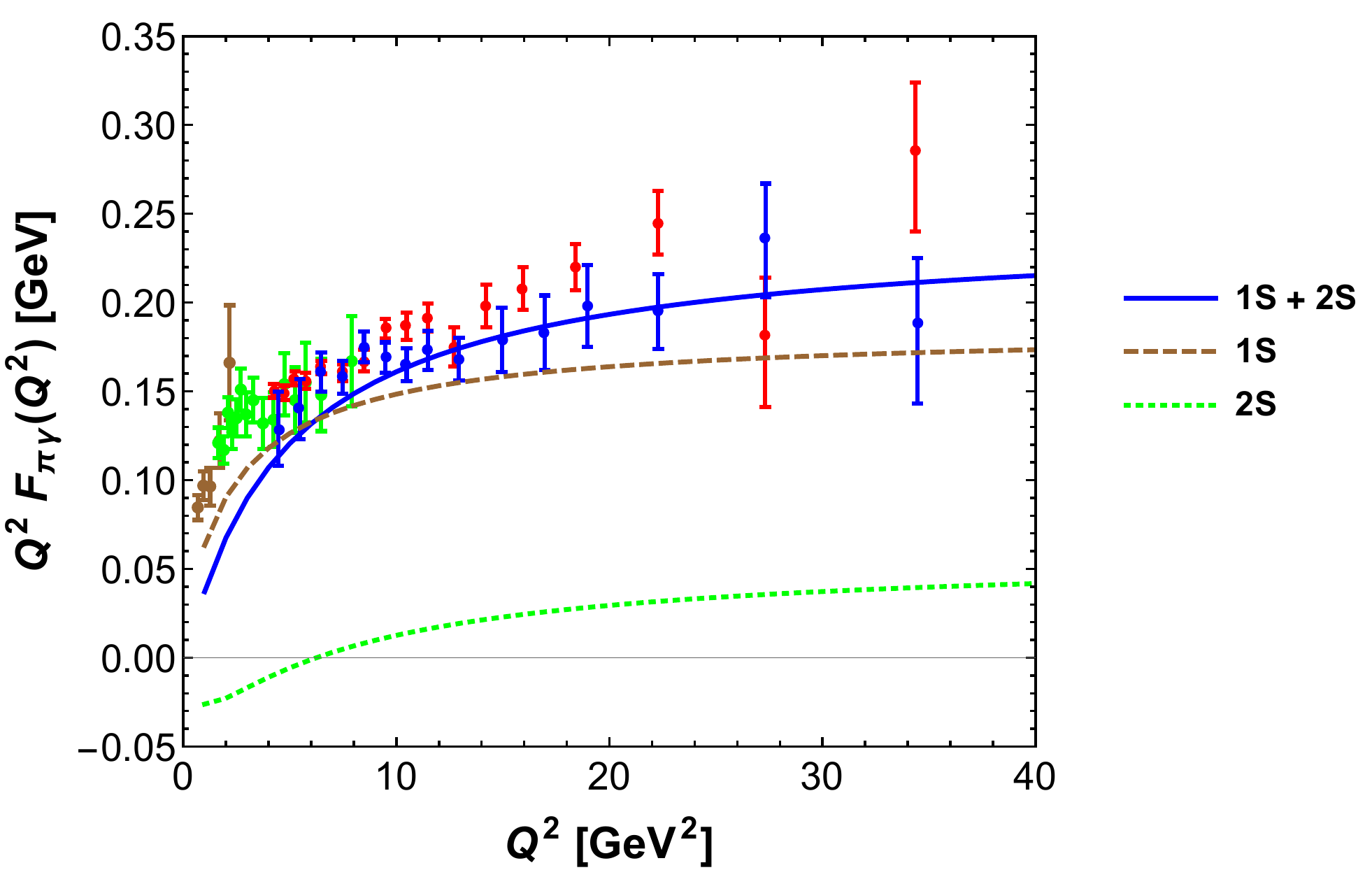}
\label{fig:Q2PionTFF}
\end{center}
\caption{\label{fig:Q2PionTFF} \small{Transition from factor shown as $Q^2F_{\pi \gamma}(Q^2)$.
The dotted and dashed curves (color online) are the contributions from the 1S and 2S states, respectively.
The solid curve represents the results when contributions from both 1S and 2S states are included. 
The results are for $P_{q\qbar}^{1S}=25\%$ and $P_{q\qbar}^{2S}=12.5\%$.
The data are taken from \cite{CELLO,CLEO,BaBar2009,Belle2012}. } }
\end{figure}

\section{Summary}

In the Fock state expansion for the light mesons, the probability of finding the $q \qbar$ pair in the 1S state is generally very small. 
Analyzing  all constraints for the pion light-cone wave function and distribution amplitude from considering the lepton and two-photon 
decays and experimental information for the quark transverse momentum, we found the probability for the 
pion to be in the 1S state is in the range of $25\% \sim 46\%$.
Apart from higher Fock states involving multiple quark-antiquark pairs and gluons,
the $q \qbar$ pair could be in the radially excited states.
We constructed the light-cone wave function for the first radially excited state (the 2S state) of the pion
employing a phenomenological connection between the light-cone wave function and the equal-time wave function in the center-of-mass frame
which is obtained using an effective simple harmonic oscillator potential.
The distribution amplitude of the 2S state at low scale has two nodes in $x$ due to the light-cone wave function having
a node in the momentum space. This is significantly different from the distribution amplitude of the 1S state which is positive 
in the whole range of $x$ at any scale. The soft QCD evolution has a much stronger impact on the distribution amplitude for the 2S state than the 1S state.

We calculate the contribution from the radially excited component (the 2S state) to the $\gamma \gamma \ra \pi^0$ transition form factor. 
It is found that this contribution grows faster (when the TFF is expressed as  $Q^2 F_{\pi\gamma}(Q^2)$) than the 1S state.
A probability for the 2S state in the range of $10\% \sim 15\%$ gives a good agreement with experimental data from the Belle Collaboration;
however, it is still very difficult to describe the fast growth shown by the data from the BaBar Collaboration.

The existence of radial excitations in the light mesons will have impacts on predictions for many observables.
A study on the TFFs for the other light mesons such as the $\eta$ and $\eta^\prime$ might provide some insights on the nature and
extent of the radial excitations in the hadrons, although the $\eta$-$\eta^\prime$ mixing might complicate the conclusions one can draw from such
a study. Such an analysis and further studies for the other processes will be presented in future work.

\section*{Acknowledgements}
I would like to thank Frank Close for inspiring discussions on the radial excitation of light mesons more than a decade ago.

\end{document}